\documentclass[final,5p,times,twocolumn]{elsarticle}
						
\usepackage{graphicx}
\usepackage{amssymb}
\usepackage{gensymb}
\usepackage{amsmath}
\usepackage{booktabs, supertabular, ctable}

\newcommand{\cm}{cm$^{-1}$}

\journal{ }

\begin{document}
\begin{frontmatter}
\author[jila]{Daniel N. Gresh\corref{cor1}}
\cortext[cor1]{Corresponding Author}
\ead{dgresh@jila.colorado.edu}
\author[jila]{Kevin C. Cossel\fnref{fn1}}
\fntext[fn1]{Current address: National Institute of Standards and Technology, 325 Broadway, Boulder, CO}
\author[jila]{Yan Zhou}
\author[jila]{Jun Ye}
\author[jila]{Eric A. Cornell}
\address[jila]{JILA, National Institute of Standards and Technology and University of Colorado, and\\
Department of Physics University of Colorado, 440 UCB, Boulder, CO 80309, USA}
\author{}
\address{}

\title{Broadband velocity modulation spectroscopy of ThF$^+$ for use in a measurement of the electron electric dipole moment}

\begin{abstract}
A number of extensions to the Standard Model of particle physics predict a permanent electric dipole moment of the electron (eEDM) in the range of the current experimental limits. Trapped ThF$^+$ will be used in a forthcoming generation of the JILA eEDM experiment. Here, we present extensive survey spectroscopy of ThF$^+$ in the 700 - 1000 nm spectral region, with the 700 - 900 nm range fully covered using frequency comb velocity modulation spectroscopy. We have determined that the ThF$^+$ electronic ground state is $X$ $^3\Delta_1$, which is the eEDM-sensitive state. In addition, we report high-precision rotational and vibrational constants for 14 ThF$^+$ electronic states, including excited states that can be used to transfer and readout population in the eEDM experiment. 
\end{abstract}

\end{frontmatter}

\section{Introduction}
\label{}

A measurement of the permanent electric dipole moment of the electron (eEDM) is a direct test of parity- and time-reversal symmetry violation \cite{Purcell1950,Hinds1997,Pospelov2005}. Electron EDM experiments have typically been performed on beams of neutral atoms \cite{Regan2002} or neutral molecules \cite{Hudson2011,Vutha2010}, with the current experimental limit of $|d_e| < 9.7\times 10^{-29}$ \textit{e}$\cdot$cm set by the ACME collaboration's molecular beam experiment \cite{Baron2014}. The JILA eEDM experiment instead uses trapped molecular ions, leveraging the large effective electric field \cite{Petrov2007,Fleig2013} and suppression of eEDM systematic errors available in molecular $^3\Delta_1$ states \cite{Kawall2004,Meyer2006}, while also achieving a long measurement coherence time \cite{Leanhardt2011}. The JILA experiment is currently using trapped HfF$^+$ \cite{Loh2013}, but ThF$^+$ is a promising candidate molecule for a future generation experiment due to its larger effective electric field \cite{Meyer2008,Denis2015,Skripnikov2015} and longer-lived $^3\Delta_1$ state. Assigning the ground state of ThF$^+$ has been experimentally challenging due to the proximity of both $^1\Sigma^+$ and $^3\Delta$ states \cite{Barker2012,Heaven2014}. Calculations suggested that $^3\Delta_1$ might be the ground state but uncertainties prevented a conclusive determination \cite{Titov2013,Denis2015}. Here, we show that the ThF$^+$ ground state is $X$ $^3\Delta_1$, and that the $a$ $^1\Sigma^+$ state lies 314 \cm\ higher in energy. This is beneficial for the eEDM experiment, since the measurement coherence time will not be limited by the spontaneous lifetime of the ``science" state. In addition to $^3\Delta_1$ and $^1\Sigma^+$, we have measured high-precision rotational and vibrational constants for 12 additional electronic states. The results reported here are benchmarks for \textit{ab initio} theory, and may assist in understanding actinide bonding. 

The lack of spectroscopic information about electronic states of ThF$^+$ above 10,000 \cm, and the difficulty of performing survey spectroscopy on a small number of trapped ions, necessitated a tool that provides broad bandwidth, high precision, high sensitivity, and ion-selectivity. Frequency-comb velocity-modulation spectroscopy (comb-vms) \cite{Sinclair2011,Cossel2012}, which combines cavity-enhanced direct frequency comb spectroscopy \cite{Thorpe2006,Adler2010} and velocity-modulation spectroscopy (vms) \cite{Stephenson2005,Crofton1983}, fulfills these requirements. With this system we have scanned the entire wavelength range of 696 - 900 nm, acquiring 3,260 \cm\ of continuous spectra at 150 MHz (0.005 \cm) step-size. Additionally, guided by the results of the comb-vms and the PFI-ZEKE spectra from Barker \textit{et al.} \cite{Barker2012}, we have performed targeted cw-laser vms scans to identify additional experimentally useful transitions and to assign electronic-state $\Omega$ values via observation of low-J rotational lines. In total, we have fit 27 rovibronic transitions in ThF$^+$ and 5 transitions in a different thoriated molecular ion, potentially ThO$^+$. We discuss the production and detection of ThF$^+$, details of the ground state assignment, and the assignment of other ThF$^+$ electronic states.

\section{Experimental details}
\label{}

\begin{figure}[h]
\includegraphics[scale=1]{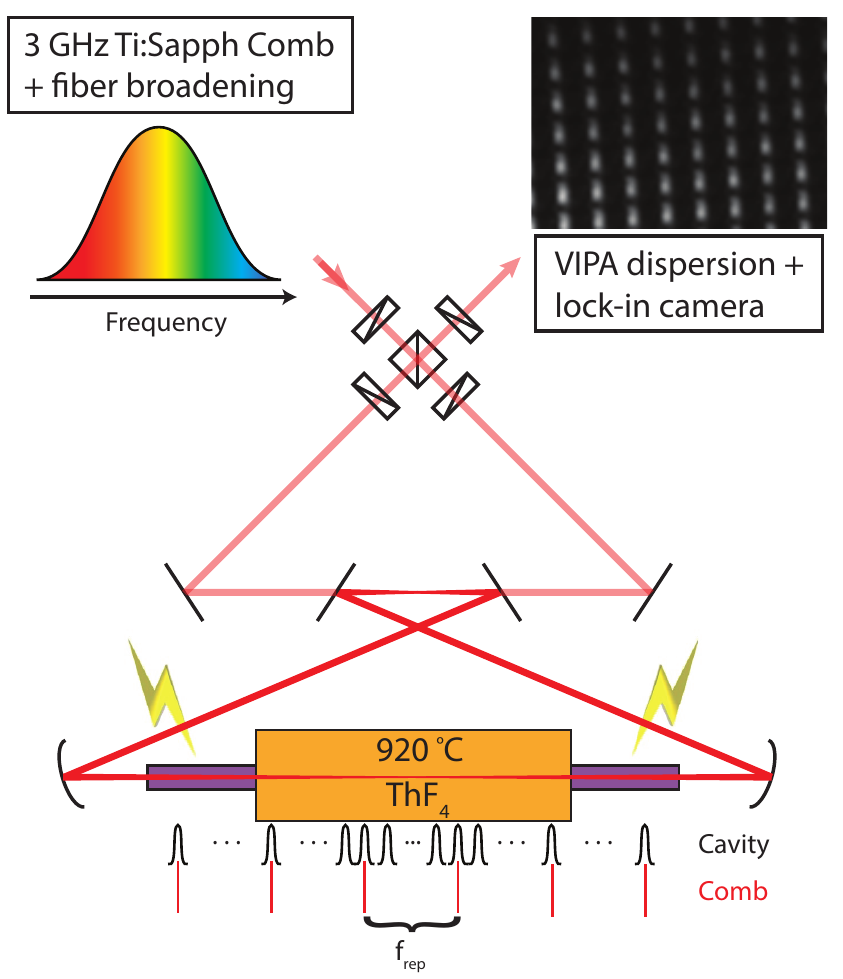}
\caption{\textbf{Experimental schematic.} A 3 GHz Ti:Sapph comb is broadened in nonlinear fiber and coupled into a bow-tie ring cavity, propagating in either direction through a discharge of ThF$_4$. Four liquid crystal variable retarders and a polarizing beam splitter control the direction of light propagation through the discharge. The cavity free spectral range is 1/25 of the comb repetition frequency. After exiting the cavity, light is sent to a cross-dispersive VIPA imaging system with single comb mode resolution and imaged onto a Heliotis lock-in camera \cite{Beer2005}. Approximately 1,500 comb modes are resolved in a single image.}
\label{fig:expdiagram}
\end{figure}

\subsection{Frequency comb vms}

Fig. \ref{fig:expdiagram} shows an overview of the comb-vms system. A 3-GHz repetition-rate titanium-doped sapphire frequency comb is spectrally broadened in nonlinear fiber and coupled into an enhancement cavity surrounding a 1-kW electrical discharge cell. A VIPA imaging system \cite{Adler2010,Nugent-Glandorf2012} resolves individual comb modes and provides 1,500 simultaneously resolved comb modes per image. 

The spectroscopically usable fundamental bandwidth of the comb is 780-850 nm, centered at 815 nm. To obtain maximum power throughput to the experiment, the output of the frequency comb is immediately coupled into the nonlinear fiber to broaden the spectrum. A half-wave plate before the nonlinear fiber input enables control of the output spectrum \cite{Coen2002}. No single fiber gave a smooth, high power spectrum in the entire 700 - 900 nm range, so we used three different fibers, replacing them as needed to cover the entire region. A Newport SCG-800-CARS fiber resulted in significant power in the 850 - 900 nm spectral region but practically no light in the 700 - 780 nm region. 85 cm of Thorlabs NL-2.8-850-02 fiber, pumped in the normal dispersion regime, provided a small amount of broadening and resulted in high power in the 730 - 780 nm range. 80 cm of Thorlabs NL-2.4-800 fiber, pumped close to the zero dispersion wavelength, resulted in most of the laser power being shifted to the 680 - 730 nm regime. 

The fiber output spectrum often contained a significant fraction of laser power shifted to a spectral region outside of the desired scanning range. After the fiber, low- and high-pass optical filters were used to select a 50 nm spectral window. The spectrum after the filters was optimized and monitored using a grating spectrometer with a USB readout. A linear polarizer was placed after the filters to eliminate the different polarization components due to nonlinear polarization rotation in the fiber. An electro-optic modulator, driven at 17 MHz, applies RF sidebands to each comb tooth.

After the modulator, the comb light is fiber-coupled to the table supporting the discharge cell. A 2.5-meter-long low-finesse (F $\approx$ 70-100) enhancement cavity surrounds the discharge tube. Every 25th cavity mode is coupled to a comb mode, and the cavity is locked to the comb using the Pound-Drever-Hall technique\cite{Drever1983}. In order to obtain maximum cavity transmission, AR-coated windows were placed on each end of the discharge apparatus (R $<$ 0.1\% per surface 700 - 900 nm), and T = 3\% cavity input couplers were used, running the cavity in the over-coupled regime. Liquid-crystal variable retarders and a polarizing beam splitter were used to rapidly switch the direction of light propagation through the cavity and the discharge tube. By subtracting the spectrum acquired from each direction, common-mode laser noise and asymmetrical discharge effects, such as an asymmetrical concentration modulation of molecules, are suppressed.

After exiting the cavity, the comb light is imaged using a 2D VIPA spectrometer \cite{Diddams2007} with single comb mode resolution onto a (Heliotis) lock-in camera \cite{Beer2005}. To fully resolve different VIPA orders, a 1800 lines/mm diffraction grating was used for the 780 - 900 nm spectral range, and a 2160 lines/mm grating was used for the 700 - 780 nm range. A cw titanium-doped sapphire laser is co-propagated with the comb light for frequency calibration. The cw laser is locked to its internal reference cavity, and the frequency is measured by a 100-MHz-accuracy wavemeter. With interleaving scans, we obtain a maximum of 150 \cm\ of continuous spectra over 1,500 simultaneous spectral channels at 150 MHz (0.005 \cm) step-size in a 25 minute total data collection time, with $3\times 10^{-8}\ \text{Hz}^{-1/2}\ \text{(spectral element)}^{-1/2}$ fractional absorption sensitivity. The acquisition time is currently limited by software and with improvements could be shortened to 10 minutes. 

The discharge tube is a straight 50 cm length of 1.7 cm inner-diameter, 1.9 cm outer-diameter quartz tubing. The tube is centered inside a home-built oven, which is heated to 920 \degree C to obtain a sufficient vapor pressure of ThF$_4$ \cite{Darnell1958,Lau1989}. Initially, smaller diameters of quartz tubing were used but were rapidly blocked during operation. The blockage is likely due to the reaction of ThF$_4$ with quartz at high temperature to form ThO$_2$ \cite{Darnell1960}, which has a low vapor pressure at 920 \degree C \cite{Ackermann1963} and should rapidly plate out along the discharge tube. A straight alumina (Al$_2$O$_3$) tube was tested but the discharge ceased to run down the length of the tube above 800 \degree C, instead opting to discharge to the grounded gas lines. 

For each oven run, approximately $1g$ of ThF$_4$ is loaded into the discharge tube, and the tube is pumped out with a scroll pump to 10 mTorr. Helium gas flows through the tube at a pressure of 5 Torr. 
Three heaters are used to maintain a temperature of 920 \degree C in the center of the oven and a temperature of 880 \degree C on either edge of the oven. Higher temperatures were found to deplete the ThF$_4$ faster but not to yield a significant increase in ThF$^+$ signal. Single oven runtimes were typically 3-4 hours before the signal fell to 80\% of its peak value due to depletion of ThF$_4$. 

During some scans we observed bands that we tentatively attribute to ThO$^+$ (see Section \ref{sec:tho}). Attempts were made to remove the ThO$^+$ signal from the discharge while monitoring ThF$^+$ and ThO$^+$ absorption strengths with cw-vms (see next section). Running the discharge with different buffer gases, such as Ne or Ar, resulted in the loss of all ThF$^+$ signal while retaining the same ThO$^+$ signal. Contrary to our expectations, seeding the discharge with SF$_6$ gas again resulted in the loss of all ThF$^+$ signal and doubled the ThO$^+$ line strengths. In all cases switching back to a He discharge resulted in an immediate return to prior ThF$^+$ line strengths. The ThF$^+$ signal was only positively affected by total oven runtime,

i.e. time when the oven was at 920 \degree C with ThF$_4$ loaded and a discharge running. After roughly 5 total hours of oven runtime, the ThO$^+$ signal decreased by a factor of two and the ThF$^+$ signal increased by a factor of two. This is likely due to Th-based discharge byproducts coating the tube walls and suppressing the ThF$_4$ \textendash\ quartz reaction. 

\subsection{cw Ti:Sapph and diode laser vms}

Twenty-one vibrational bands were fit in the comb-vms spectrum. The total spectral coverage of the comb-vms system is limited by coatings on the VIPA etalon and by the decreased quantum efficiency of the silicon-based lock-in camera at wavelengths longer than 900 nm. Cw-laser vms was used to extend spectral coverage beyond the 700 - 900 nm range in order to identify additional favorable electronic states for coherent population transfer between  $X$-$a$ and to increase the signal-to-noise ratio on low-J rotational lines. For example, low-J lines of the $\Omega = 0^-\!\leftarrow\!^3\Delta_1$ (0, 0) band were observed at 685.4 nm, a wavelength inaccessible with the current optics coatings in the comb-vms system (see Section \ref{sec:3d1}). 

Due to the lower complexity of cw scans, a higher single-pass signal-to-noise ratio can be obtained. However, the total data acquisition time is more than 20 times slower per bandwidth than comb-vms. For the cw scans, the enhancement cavity was removed, and the laser was split into two components with a power ratio of approximately 2:1 that counterpropagated through the discharge tube. Roughly 1 mW of power is contained in the lower power beam. The lock-in signal is detected on the difference port of an autobalancing detector \cite{Hobbs1997}. The best scan rate achieved is 1 \cm\ in 4 minutes with a 300 ms lock-in time constant, limited by reading the wavemeter and scanning slowly enough to maintain laser lock. The best single-pass fractional sensitivity reached is $2\times 10^{-8}\ \text{Hz}^{-1/2}$, about a factor of 2 above shot noise at 1 mW, limited by a combination of electrical and acoustic pickup from the discharge. 

Low-J lines of the $\Omega = 0^-\!\leftarrow\!^3\Delta_1$ (0, 0) band at 685.4 nm were scanned using an external-cavity diode-laser with a 680 nm diode. The high-precision rotational and vibrational constants measured from the $\Omega = 0^-\!\leftarrow\!^3\Delta_1$ (0, 1), (1, 2), (2, 3), (0, 2), and (1, 3) bands allowed us to pinpoint line centers in the (0, 0) vibrational band to 0.1 \cm\ accuracy. Feed forward to the diode current resulted in a mode-hop free scan range of 0.2 \cm. Fractional sensitivity with 50 $\mu$W optical power was limited by electrical pickup to $1\times 10^{-7}\ \text{Hz}^{-1/2}$, yielding signal-to-noise $> 10$ on the weakest low-J lines with a 1 s lock-in time constant. 

\section{Results and analysis}
\label{}

\begin{figure*}[t]
\includegraphics[scale=0.97]{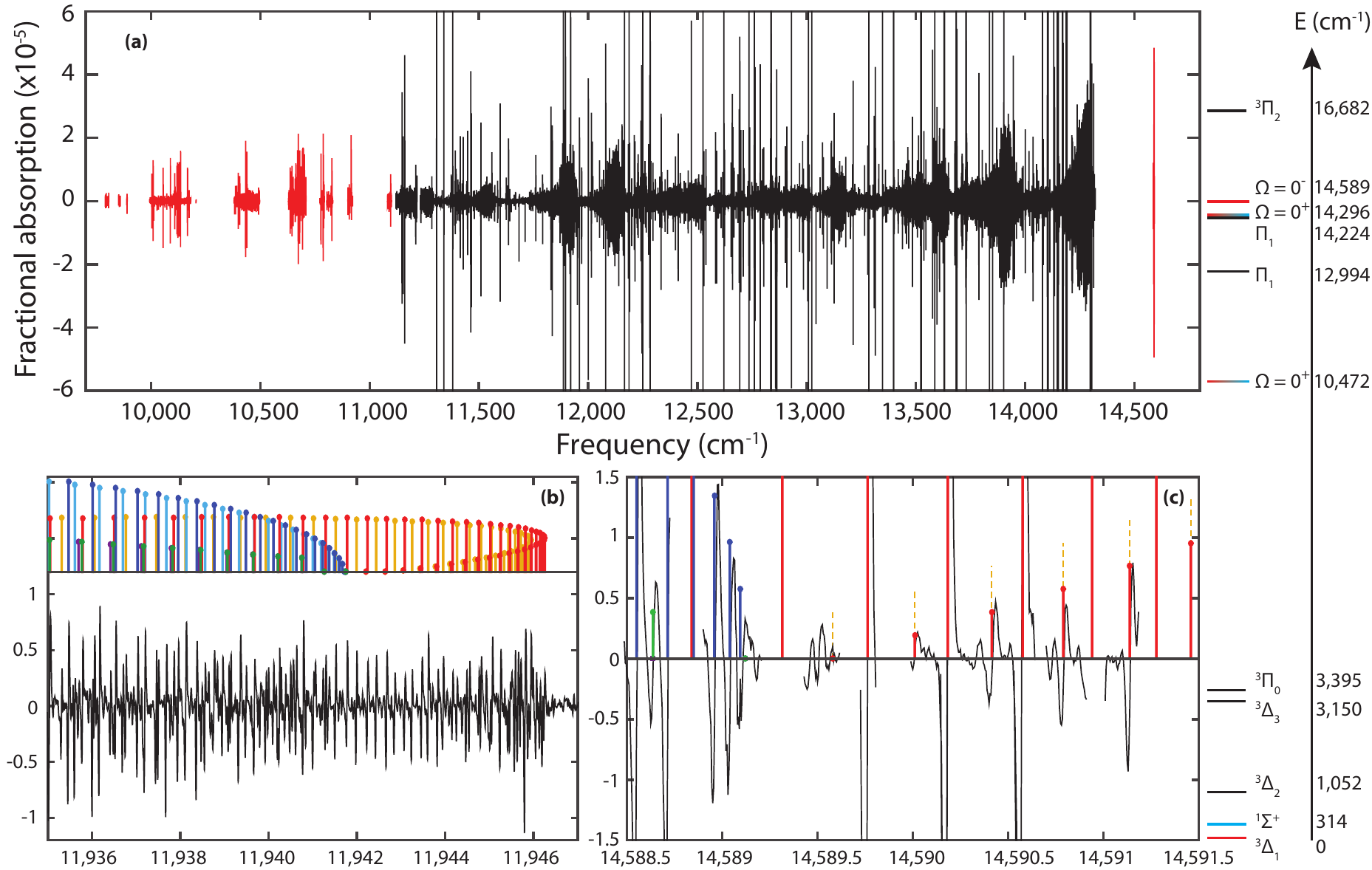}
\caption{\textbf{All acquired ThF$^+$ spectrum.} The energy level diagram on the right indicates all experimentally observed electronic energy levels. All y-axes are in 10$^{-5}$ single pass fractional absorption. All x-axes are in \cm. (a) All ThF$^+$ spectra acquired with comb-vms (black) and cw-vms (red). (b) Fitted positions to $\Pi_1 [12.99] \leftarrow\ ^3\Delta_2$ (0, 0). Note the presence of six branches -- two P- (purple and green), two Q- (dark blue and light blue) and two R- (red and orange) -- due to $\Lambda$-doubling in each electronic state. (c) Low-J lines in the $\Omega = 0^-\!\leftarrow\!^3\Delta_1$ (0, 0) band. The solid colors indicate different branches for an $\Omega' = 0 \leftarrow \Omega'' = 1$ transition (P- green, Q- blue, R- red). The dashed orange lines represent the R-branch intensity if the transition was instead $\Omega' = 1 \leftarrow \Omega'' = 0$. In each case the overall band intensity is set by scaling the line intensities to match the strongest R-branch lines, which are off-scale peaks marked by solid red lines. P(1) is present, and R(0) is clearly absent.}
\label{alldata}
\end{figure*}

We have observed 14 different ThF$^+$ electronic states. To indicate state assignments Hund's case (a) notation is used when possible, and case (c) notation is used when only the $\Omega$-value is known. All of the acquired data is shown in Fig. \ref{alldata} (a): a total of 3,260 \cm\ was acquired with comb-vms (black), and 400 \cm\ with cw-vms (red). ThF$^+$ bands often presented with at least a single P-, Q-, and R-branch, and frequently had two of each branch due to $\Lambda$-doubling in both electronic states. Fig. \ref{alldata} (b) shows a typical six branch vibrational band, with a clear R-branch band head and a strong Q-branch. The $\Lambda$-doubling is strong enough to fully resolve the two parity components by J = 3. 

For an observed vibrational band, rotational constants are determined by fitting assigned line centers, defined as the zero-crossing, to a general energy expression:
\begin{align}
\label{eqn:hamiltonian}
\begin{split}
E =&\ \nu_0 + (B' - s'k'/2)J'(J'+1)-(D'-s'k'_D/2)J'^2(J'+1)^2 \\
&- [(B'' - s''k''/2)J''(J''+1) -(D''-s''k''_D/2)J''^2(J''+1)^2].
\end{split}
\end{align}
Here, $s$ is a parity term: $s = +1$ for \textit{e}-symmetry and $s = -1$ for \textit{f}-symmetry, and $k$ and $k_D$ are generic $\Lambda$-doubling constants proportional to J(J+1) and J$^2$(J+1)$^2$, respectively. Bands with $\Lambda$-doubling observed in each electronic state have all six branches fit simultaneously, with $s = \pm 1$ corresponding to the two parity levels in each state. Most observed transitions have the prominent $\Lambda$-doubling contribution arising from the upper state; in these transitions $k''$ and $k''_D$ are set to 0. 

Vibrational constants are determined from the relationship,

\begin{align}
\label{eqn:vibration}
\begin{split}
\nu_{v'v''} = T'_e - T''_e &+ \omega'_e(v' + 1/2) - \omega_e\chi'_e(v' + 1/2)^2 \\
&- [\omega''_e(v' + 1/2) - \omega_e\chi''_e(v' + 1/2)^2],
\end{split}
\end{align}
where $\nu_{v'v''}$ is the $(v', v'')$ band origin, $\omega_e$ is the vibrational constant, $\omega_e\chi_e$ is the anharmonicity constant, and T$_e$ is the energy to the minimum of the electronic potential. For states with too few transitions observed to directly determine $\omega_e$ and $\omega_e\chi_e$, a Morse potential was assumed, allowing the substitution

\begin{equation}
\omega_e\chi_e = \frac{\alpha^2_e\omega^2_e}{36B^3_e} + \frac{\alpha_e\omega_e}{3B_e} + B_e,
\end{equation}
where $\alpha_e$ is the rotational-vibrational coupling constant and $B_e$ is the equilibrium rotational constant.

Vibrational assignments are determined by comparing observed band intensities to Franck-Condon factors and a thermal vibrational distribution. With prior knowledge of vibrational constants from Barker \textit{et al.} \cite{Barker2012}, and our measured $B$ values, Franck-Condon factors were calculated and compared to observed intensities for the initial bands measured. With the rotational and vibrational constants we have observed in ThF$^+$, the vibrational (0, 0) is always the strongest band. The highest intensity bands in the $\Delta v$ = $\pm 1$ vibrational series are about 1.5-2 times weaker than the (0, 0) band. 

Due to the high density of electronic states, the relatively small difference in vibrational constants for different electronic states, and the high oven temperature, the ThF$^+$ spectrum is often very congested. Fits are performed by manually identifying branches and assigning line centers to the rotational structure given by Eqn. \ref{eqn:hamiltonian}. 

We have developed a simple automatic line finder to assist with identifying bands in the most congested regions. Local maxima are identified and then matched with the nearest zero crossing of the correct line-shape phase within the observed linewidth of 1 GHz. A minimum intensity threshold is specified to avoid assigning noise to real lines, and a maximum intensity threshold is used to avoid assigning strong atomic features to molecular rotational lines. 

The identified lines are plotted in a Loomis-Wood style figure, which displays J = 0-100 residuals on a single figure. Over such a large scale, pattern-forming rotational sequences can be easier to identify. By modifying $B$ values, $D$ values, and the band origin, $\nu_0$, it is possible to precisely identify the line centers of hundreds of rotational lines at once. Additionally, the large J-range of the plots enables clear identification of smaller high-J effects, such as $\Lambda$-doubling scaling as J$^4$, which typically manifests in the spectrum as lines splitting into doublets above J = 40. 

\subsection{$^3\Delta_1$ ground state assignment}
\label{sec:3d1}

\begin{figure*}[t]
\includegraphics[scale=1]{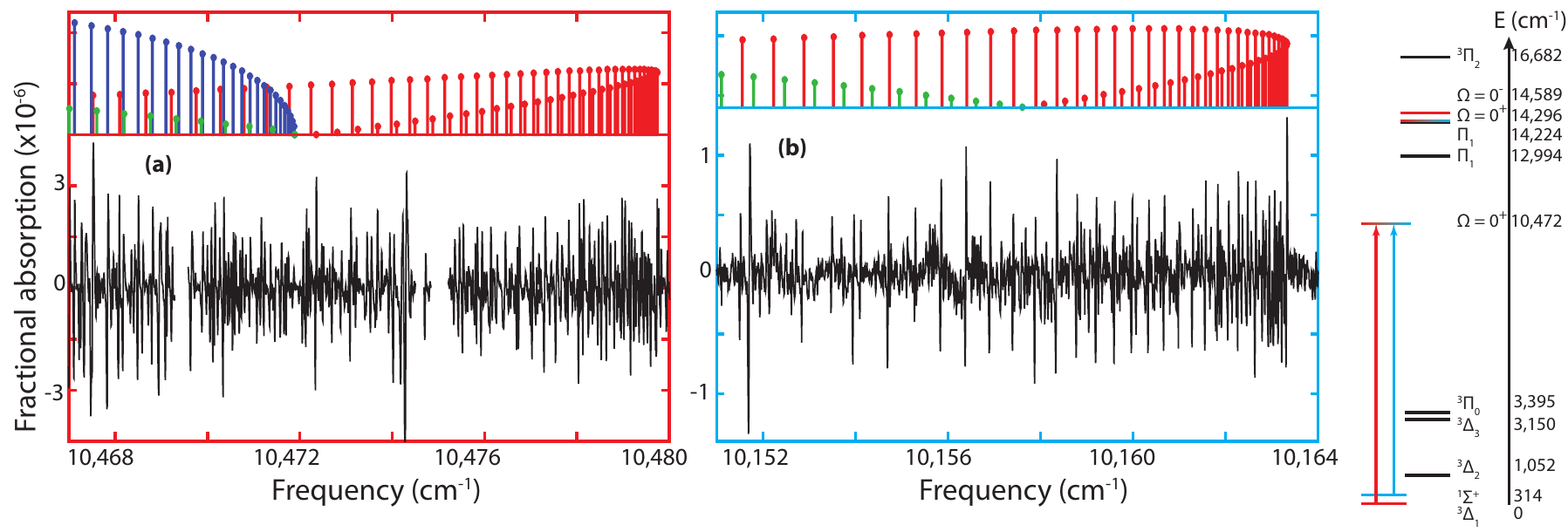}
\caption{\textbf{Spectrum of $^3\Delta_1$ and $^1\Sigma^+$ coupling to a common $\Omega = 0^+$ state.} (a) An $\Omega = 0^+ \leftarrow\ ^3\Delta_1$ vibrational band is observed at $\nu_0 = 10471.889(7)$ \cm\ (red axes). A strong Q-branch (blue markers) is observed in addition to the P- (green markers) and R-branches (red markers). (b) A transition is observed 314 \cm\ lower in energy at $\nu_0 = 10157.607(7)$ \cm\ (blue axes). This transition has no Q-branch, indicating $\Omega'' = \Omega' = 0$. The transition ordering indicates that $^3\Delta_1$ is lower in energy as illustrated in the energy level diagram on the right.}
\label{commonspec}
\end{figure*}

Assignment of the low-lying $^3\Delta_1$ state is of paramount importance to the eEDM experiment. Initial PFI-ZEKE experiments from Barker \textit{et al.} \cite{Barker2012} observed both the $^1\Sigma^+$ and $^3\Delta_1$ states and measured the electronic energy difference to be $315(1)$ \cm. However, due to the inability to fully resolve low-J lines, compounded by the similarity of $B$ and $\omega_e$ values for the $X$ and $a$ states, it was not possible to conclusively assign either as the ThF$^+$ ground state \cite{Barker2012}. Our assignment necessitates observation of an excited electronic state with coupling to both the $^1\Sigma^+$ and $^3\Delta_1$ states. Such a state will have a spectral signature of two vibrational bands with band origins separated by the $X$-$a$ energy difference. In order to couple to both lower states, the upper state must be either $\Omega = 0^+$ or $\Omega = 1$. The final assignment uses measurement of the $\Lambda$-doubling and an observation of low-J lines. 

In the comb-vms data we have fit transitions at $\nu_0$ = 13935.996(7) \cm\ and $\nu_0$ = 13286.716(7) \cm\ with one set of P-, Q-, and R-branches and a nonzero $k$ constant. The presence of a nonzero $k$ and one of each branch implies that the two electronic states in the transition must be $\Omega$ = 0 and $\Omega$ = 1. The small magnitude of $k$, $|k|$ = 0.869(7)$\times 10^{-4}$ \cm, makes it impossible to differentiate between $k$ in the upper state, $k'$, or in the lower state, $k''$. Because $\omega_e$ for the $^1\Sigma^+$ and $^3\Delta_1$ states is so similar it could not be used for electronic state identification. Based on calculated Franck-Condon factors and the intensity of the three transitions observed at $\nu_0$ = 13935.996(7), 13876.539(8), and 13817.279(8) \cm, and the two transitions observed at $\nu_0$ = 13286.716(7) and 13231.099(8) \cm, the transition at $\nu_0$ = 13935.996(7) \cm\ was identified as the $\Omega = 0^-\!\leftarrow\!^3\Delta_1$ (0, 1) vibrational band. 

The lower state rotational constant, $B_0 = 0.24264(3)$ \cm, and the vibrational constant, $\omega_e = 656.96(1)$ \cm, both agree with Barker \textit{et al.} \cite{Barker2012} for low-lying states of ThF$^+$. Also, the J$^2$ $\Lambda$-doubling proportionality constant, $|k| = 0.869(7) \times 10^{-4}$ \cm, is consistent with expectations for a $\Delta_{\Omega = 1}$ state: J$^2$ scaling and small magnitude \cite{Brown1987}. 
Thus, we tentatively assign the lower state to $^3\Delta_1$. 

To confirm the lower state $\Omega = 1$ assignment we measured low-J lines of the $\Omega = 0^-\!\leftarrow\!^3\Delta_1$ (0, 0) band at $\nu_0$ = 14589.09(2) \cm, as shown in Fig. \ref{alldata} (c). Because of the high oven temperature the spectrum is typically congested, but we were able to measure low-J lines of this band due to the high line strength. Higher-lying Q- and R-branch lines were scanned to confirm rotational constants and to fix the overall intensity of the band. Two compelling pieces of evidence are seen. First, the R(0) line, indicated by the dashed orange line, is absent, while the P(1) line, indicated by the solid green marker, is present. We clearly see R(1)-R(4). P(2)-P(4) are covered by strong Q-branch lines. Secondly, the H{\"o}nl-London factors \cite{Herzberg1950} for an $\Omega'' = 0 - \Omega' = 1$ transition considerably increase the intensity of the low-J R-branch lines relative to an $\Omega'' = 1 - \Omega' = 0$ transition, indicated by the dashed orange lines and the solid red lines, respectively. The observed R-branch line intensities agree with $\Omega'' = 1$. This confirms that the lower state is $\Omega = 1$ and the upper state is $\Omega = 0$. 

We have observed another transition at $\nu_0$ = 10471.889(7) \cm, with one of each branch and nonzero $k$, shown in Fig. \ref{commonspec} (a). The fitted values of $B$ and $D$ agree with $B_0$ and $D_0$ for $^3\Delta_1$. Additionally, the magnitude of $k$ agrees with the previously measured transition at $\nu_0$ = 13935.996(7) \cm\ but the sign is changed, which is expected if the upper state is of a different overall parity, e.g. an $\Omega = 0^+$ instead of an $\Omega = 0^-$ state. Therefore, the lower state in this transition is also $^3\Delta_1$. 

A vibrational band with no Q-branch is observed 314.282(7) \cm\ lower in energy, consistent with the  $X$-$a$ separation measured by Barker \textit{et al.} \cite{Barker2012}, with $\nu_0$ = 10157.607(7) \cm. The band origin is shown in Fig. \ref{commonspec} (b). The absence of a Q-branch indicates the two electronic states in the transition both have $\Omega$ = 0, giving strong evidence for the lower state in this transition being $^1\Sigma^+$. To confirm the ground state assignment, the upper electronic state must be the same as the one observed in the $\nu_0$ = 10471.889(7) \cm\ transition. 

\begin{figure}[h]
\includegraphics[scale=1]{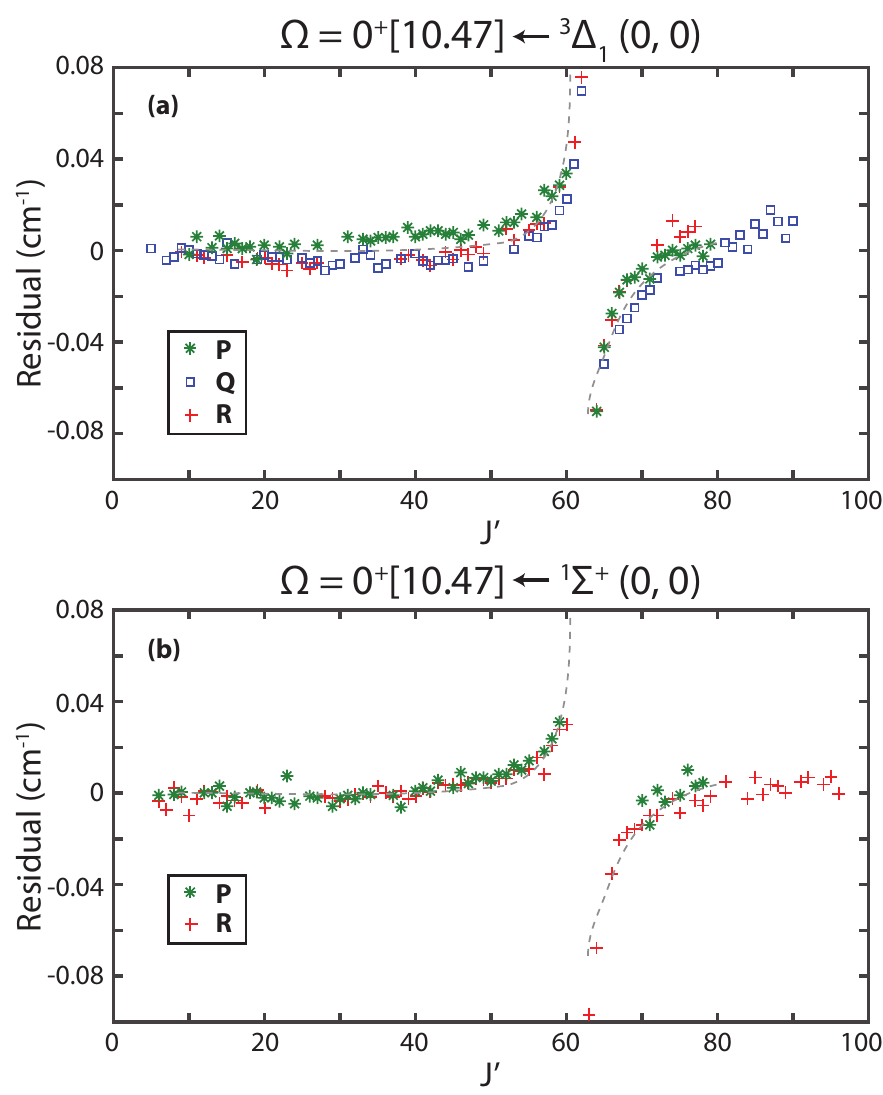}
\caption{\textbf{Fit residuals for $^3\Delta_1$ and $^1\Sigma^+$ coupling to a common upper state.} The fit residuals of each transition exhibit an avoided crossing perturbation of the same magnitude at the same rotational quantum number, confirming the upper electronic state is the same in each transition. The dashed grey lines on each figure are guides to the eye and are identical. (a) Fit residuals for $\Omega = 0^+ [10.47]\!\leftarrow\!^3\Delta_1$ (0, 0). (b) Fit residuals for $\Omega = 0^+ [10.47]\!\leftarrow\!^1\Sigma^+$ (0, 0).}
\label{fig:residuals}
\end{figure}

With no Q-branch the band cannot be fit \textit{a priori}. However, by fixing $B'$ and $D'$ to the values measured in the $\nu_0$ = 10471.889(7) \cm\ transition, we obtained a good fit with reasonable residuals. The lower-state rotational constant fits to $B'' = 0.24542(3)$ \cm. This is larger than the fit $B(^3\Delta_1)$, as intuitively expected since an $s^2$ electron configuration (in $^1\Sigma^+$) is typically more tightly bound than an $sd$ configuration (also observed in HfF$^+$ \cite{Cossel2012}). 

The fit residuals provide the final piece of compelling evidence for the upper electronic state being the same. Fit residuals for the $\Omega = 0^+ [10.47]\!\leftarrow\!^3\Delta_1$ (0, 0) at $\nu_0$ = 10471.889(7) \cm\ are plotted in Fig. \ref{fig:residuals} (a). A strong avoided crossing perturbation is evident at approximately $J'$ = 62. Other $^3\Delta_1$ fits have no avoided crossing, so the perturbation must be present in the $\Omega = 0^+$ upper electronic state. Fit residuals for the $\nu_0$ = 10157.607(7) \cm\ transition, shown in Fig. \ref{fig:residuals} (b), display an avoided crossing of the same magnitude again at $J'$ = 62. This confirms that the upper state is the same for the two transitions, and thus that the assignments of $^1\Sigma^+$ and $^3\Delta_1$ are correct. Since the $\Omega = 0^+ [10.47]\!\leftarrow\!^1\Sigma^+$ (0, 0) transition frequency is lower than that of the $\Omega = 0^+ [10.47]\!\leftarrow\!^3\Delta_1$ (0, 0) transition, $^3\Delta_1$ is the electronic ground state. 

\subsection{Additional ThF$^+$ electronic states}

In addition to the $^1\Sigma^+$ and $^3\Delta_1$ electronic states we have observed rovibronic transitions that we have assigned to originate from the $^3\Delta_2$ and $^3\Delta_3$ states. These are given in Table \ref{transitions} as the $\Pi_1 [12.99]\!\leftarrow\!^3\Delta_2$, the $\Pi_1 [14.22]\!\leftarrow\!^3\Delta_2$, and the $^3\Pi_2\!\leftarrow\!^3\Delta_3$ groups of transitions. 

For the $^3\Delta_2$ state, five vibrational bands of $\Pi_1 [12.99]\!\leftarrow\!^3\Delta_2$ and one band of $\Pi_1 [14.22]\!\leftarrow\!^3\Delta_2$ were observed. Each band has two of each P-, Q-, and R-branch originating from a common band origin, which indicates that $\Lambda$-doubling is present in each state. A small section of the $\Pi_1 [12.99]\!\leftarrow\!^3\Delta_2$ (0, 0) band is shown in Fig. \ref{alldata} (b). Two strong Q-branches are seen, indicating that $\Omega' \neq \Omega''$. The $\Lambda$-doubling has a strong J$'^2$ dependence, consistent with the upper state being predominantly $\Pi_{\Omega = 1}$ \cite{Brown1979,Brown1987}. $B''_e = 0.24342(2)$, similar to the value of $B_e(^3\Delta_1) = 0.24311(7)$, and consistent with the measured value for $^3\Delta_2$ from Barker \textit{et al.} \cite{Barker2012}. Additionally, $\omega''_e$ is consistent with the value measured by Barker \textit{et al.} and again similar to $\omega_e(^3\Delta_1)$. 

To confirm this assignment, we observed low-J lines in the $\Pi_1 [12.99]\!\leftarrow\!^3\Delta_2$ (0, 1), (0, 0), and (1, 0) bands as well as in the $\Pi_1 [14.22]\!\leftarrow\!^3\Delta_2$ (0, 0) band using the cw Ti:Sapph. In each case, Q(1) is absent and Q(2) is present, thus leaving $\Omega'' = 1$ or $2$ as the only possibilities. R(1) is absent in each band, and the intensity of the low-J P- and R-branch lines is consistent with $\Omega'' > \Omega'$. Therefore, $\Omega'' = 2$, $\Omega' = 1$, and the lower electronic state is $^3\Delta_2$. 

The assignment for $^3\Delta_3$ is similar. Three vibrational bands of $^3\Pi_2\!\leftarrow\!^3\Delta_3$ were observed, with each band again having two of each P-, Q-, and R-branch at a common band origin. Here though the $\Lambda$-doubling has a strong J$'^4$ dependence, as expected for a $^3\Pi_{\Omega = 2}$ upper state \cite{Brown1979}. $B''_e = 0.24388(7)$ is similar to that of $^3\Delta_1$ and $^3\Delta_2$, and consistent with Barker \textit{et al.} \cite{Barker2012}. Scanning near the band origins of the $^3\Pi_2\!\leftarrow\!^3\Delta_3$ (0, 1) and (0, 0) bands with the cw laser revealed that P(2) is absent and P(3) is present. Low-J Q-branch lines could not be resolved due to contamination from other bands. However, the intensity of the low-J P- and R-branch lines is again consistent with $\Omega'' > \Omega'$, concluding that $\Omega'' = 3$ and $\Omega' = 2$. This confirms the lower state $^3\Delta_3$ assignment. The assignment of upper state $^3\Pi_2$ is based on the strong, $J'^4$ $\Lambda$-doubling, and the expected strong transition strength for a $^3\Pi_2 \leftarrow\ ^3\Delta_3$ transition. 

Several bands exhibit strong perturbations at high-J, which manifest as fit residuals that deviate from Eqn. \ref{eqn:hamiltonian} at $J \gtrsim 60$. To obtain uniform fit residuals at high-J it was necessary to limit the range of J-values, or to include additional fit parameters in the Hamiltonian, such as $HJ^3(J+1)^3$, or in some cases even higher powers of J. Higher vibrational levels of the $\Pi_1 [12.99]$ state appear to become perturbed, likely as the result of proximity to a perturbing state near 14,000 \cm. This is evident by three different observations. First, $k_D'$ has a strong vibrational dependence, increasing from $2.1(2)\times 10^{-9}$ \cm\ in the $v' = 1$ vibrational state, to $7.7(8)\times 10^{-9}$ \cm\ in the $v' = 2$, and to $10.1(7)\times 10^{-9}$ \cm\ in the $v' = 3$, the highest vibrational state we observe. Secondly, fit residuals deviate from Eqn. \ref{eqn:hamiltonian} at J $>$ 80 for the $v' = 2$ and $v' = 3$ vibrational states, but not for the $v' = 0$ and $v' = 1$. Finally, assuming a Morse potential, $\omega_e$ and $\omega_e\chi_e$ can be calculated for both the lower and upper electronic states using the (0, 1), (0, 0), and (1, 0) bands. However, the origins of the (2, 1) and (3, 2) bands are inconsistent with the values calculated, indicating that the upper state electronic potential deviates from the Morse approximation beyond $v' = 1$. Similarly, $k'_D$ is strongly vibrationally dependent in the $^3\Pi_2\!\leftarrow\!^3\Delta_3$ group of transitions, decreasing from $74(4)\times 10^{-9}$ \cm\ in the $v' = 0$, to $41(9)\times 10^{-9}$ \cm\ in the $v' = 1$. This is likely due to a high density of unobserved (dark) electronic states. 

Perturbations are also evident in a number of other vibrational bands. The $\Omega = 0^+ [14.29]\!\leftarrow\!^3\Delta_1$ (0, 0) band shows a strong deviation from the expression of Eqn. \ref{eqn:hamiltonian} above $J = 50$. The (0, 1) band has the same structure in the fit residuals. The (1, 1) band, which ordinarily would be strong, is not seen, and the (2, 2) band is observed at $\nu_0 =$ 13639.836(4) \cm, only a difference of 2 \cm\ from the band origin of the (0, 1). This deviation from typical vibrational spacings indicates that the upper state electronic potential is heavily perturbed. Similarly, the $\Pi_1 [14.22]\!\leftarrow\!^3\Delta_2$ (0, 0) band fit residuals deviate from the standard form above $J = 50$. Here, no other vibrational levels have been observed; the (0, 1) band is obscured by the strong $\Pi_1 [12.99]\!\leftarrow\!^3\Delta_2$ (1, 0), and the (1, 1) and (2, 2) are not observed. 

Four additional ThF$^+$ electronic states were observed but not conclusively assigned. In all cases, the lower electronic state $B$ values are not consistent with the states observed by Barker \textit{et al.} \cite{Barker2012}. This suggests that T$_e$ $>$ 4,000 \cm, which in turn suggests that this population is strongly superthermal. Franck-Condon factors indicate that the band at $\nu_0 =$ 10702.830(4) \cm\ is a vibrational (0, 0) and the two bands at $\nu_0 =$ 10085.683(7) and 10050.200(7) \cm\ are the (0, 1) and (1, 2), respectively. A third band was observed at $\nu_0 \approx$ 10,014 \cm, consistent with the location and intensity of the (2, 3), but was not fit. Each transition has two of each P-, Q-, and R-branch, again indicating $\Lambda$-doubling in each electronic state. The lower electronic state exhibits $J''^4$-scaling $\Lambda$-doubling with a strong vibrational dependence: $k''_D = 34.0(7)\times 10^{-9}$ \cm\ in the $v'' = 0$, and it decreases to $14(1)\times 10^{-9}$ \cm\ in the $v'' = 1$ and $5.5(3)\times 10^{-9}$ \cm\ in the $v'' = 2$. Low-J line intensities in the (0, 0) and (0, 1) bands reveal that $\Omega'' < \Omega'$. R(1) is absent and R(2) is present. The magnitude and scaling of the $\Lambda$-doubling, the lower value of $B$, and the low-J lines suggest that the lower electronic state may be a $^3\Pi_2$ state. With the vibrational assignment we can calculate constants for the electronic states. For the lower state, we obtain $B''_e = 0.23402(5)$ \cm, $\omega''_e = 619.8(2)$ \cm, $\omega_e\chi''_e = 1.34(9)$ \cm, and $\alpha''_e = 0.74(4)\times 10^{-4}$ \cm. For the upper state we measure: $B'_e = 0.22363(5)$ \cm, $\omega'_e = 593.1(4)$ \cm, $\omega_e\chi'_e = 1.70(9)$ \cm, and $\alpha'_e = 0.89(4)\times 10^{-4}$ \cm.

A pair of transitions was observed at $\nu_0 =$ 12721.532(7) \cm\ and $\nu_0 =$ 12661.053(7) \cm. Each band has a single P- and R- branch and a strong Q-branch. $k$ and $k_D$ both fit to 0, indicating very small $\Lambda$-doubling in one or both of the states. Due to spectral congestion and the relatively low intensity of the band, no low-J lines were observed. The intensity of the bands is consistent with vibrational quantum numbers of $\Delta v = +1$ or $\Delta v = -1$. A third band, likely the $(v'+2, v''+2)$, can be barely resolved at $\nu_0 \approx$ 12,600 \cm, but it is too weak to fit. The (0, 0) is typically the strongest band, but is not immediately observed when searching within a reasonable range of $\omega_e$ values; it is likely obscured by the presence of stronger bands in these regions. Without a definitive vibrational assignment we cannot conclusively determine $B_e$, $\omega_e$, or $\omega_e\chi_e$. 

\subsection{ThO$^+$}
\label{sec:tho}

We have also fit five bands that do not originate from ThF$^+$. Based on the rotational constants, these bands appear to be originating from two different electronic states, separated by about 582 \cm, coupling to a common upper state. The fit rotational constants are nearly 1.5 times larger than those for ThF$^+$ values: $B \approx 0.34$ \cm, compared to $B(\text{ThF}^+) \approx 0.24$ \cm. Additionally, the bands must be fit with half-integer J, indicative of a molecule with a spin doublet. ThO$^+$, experimentally studied by Goncharov \textit{et al.} in 2006 \cite{Goncharov2006}, seems a promising candidate. However, our measured constants, given in Table \ref{tho}, do not form a fully consistent picture with the values measured by Goncharov \textit{et al.}, so we warn that our identification of ThO$^+$ as the unknown species is tentative. 

ThO$^+$ bands were fit to the form given in Eqn. \ref{eqn:hamiltonian}, except with J half-integer. All five transitions fit well out to J = 100 with no $\Lambda$-doubling terms, \textit{i.e.} $k = k_D = 0$. The vibrational assignment is tentative and is based on overall band intensities as predicted by Franck-Condon factors. The electronic state with $B_0 = 0.33984(2)$ \cm\ is consistent with the $^2\Delta_{3/2}$ state of ThO$^+$. $B' = 0.32610(1)$ \cm\ is a reasonable value for a higher-lying electronic state. However, the third electronic state, with $B_0 = 0.33900(2)$ \cm, is only consistent with a higher-lying $^2\Sigma^+$ or $^2\Delta_{5/2}$ vibrational level. The upper electronic state in each group of transitions appears to be identical; $B$ is consistent to within 0.00002 \cm. If the upper state is the same, the two lower states must be separated in energy by 12146.812(1) \cm\ - 11564.329(1) \cm\ = 582.483(2) \cm. However, no two measured states of ThO$^+$ \cite{Goncharov2006} are separated by this energy. Because these states had not been observed by Goncharov \textit{et al.} \cite{Goncharov2006}, it is possible that the ThO$^+$ population distribution in the discharge is superthermal, resulting in the observation of high-lying electronic states. Finally, we observed a ThO$^+$ characteristic band head near 9800 \cm, but it was sufficiently out of range of our cw Ti:Sapph that we did not scan enough of the band to attempt a fit. 

\section{Conclusion}
\label{}

We have determined that $X$ $^3\Delta_1$ is the electronic ground state of ThF$^+$, and $a$ $^1\Sigma^+$ is the first excited state, located 314 \cm\ above the ground state. This was assigned by observing two $\Omega = 0^+$ states that couple to both $X$ and $a$, at T$_e$ = 10,472 and 14,296 \cm. In an eEDM experiment it may be beneficial to first selectively ionize neutral ThF into the ThF$^+$ $^1\Sigma^+$ state, then transfer population to the ground state to obtain a pure population in a single Zeeman, Stark, and hyperfine sublevel \cite{Leanhardt2011}. Either $\Omega = 0^+$ state can be used to transfer population between the two states. Transitions to the $\Omega = 0^+ [14.29]$ state are nearly a factor of 10 stronger in relative intensity than to the $\Omega = 0^+ [10.47]$ state; however, the weaker transition is notably on par with the transition strength to the $^3\Pi_{0+}$ state observed in HfF$^+$, which is already strong enough for coherent population transfer \cite{Cossel2012}. 

For the eEDM experiment, the $\Lambda$-doubling in $^3\Delta_1$ is important for setting the voltage required to fully polarize the molecule. The $\Lambda$-doubling in the ThF$^+$ $^3\Delta_1$ state is about 7 times larger than in the HfF$^+$ $^3\Delta_1$ state \cite{Cossel2012}. The larger value can be explained by the higher density of states at low energy. The $^3\Delta_1$ $\Lambda$-doubling can arise via L-uncoupling with a nearby $^3\Pi_0$ state, which can in turn couple to a $\Sigma$-state via a spin-orbit interaction \cite{Field2015,Lefebvre-Brion1986}. Barker \textit{et al.} observed a $^3\Pi_0$ state at E = 3,395 \cm\ \cite{Barker2012}, three times closer in energy to $^3\Delta_1$ than it is in HfF$^+$. This would in turn lower the energy of $^3\Pi_1$, another potential interaction pathway, relative to HfF$^+$. Additionally, it is not unreasonable to expect a lower-lying $\Sigma$ state to account for the remaining difference. 

In addition, we have measured and assigned rotational and vibrational constants of the $^3\Delta_2$ and $^3\Delta_3$ states, and measured rotational constants, vibrational constants, and electronic energies for several intermediate electronic states. These values will provide a benchmark for \textit{ab initio} calculations. Derived constants for ThF$^+$ electronic states are given in Table \ref{stateconstants}. A comparison to other ThF$^+$ experimental and theory results is given in Table \ref{theorycomp}. Other than ambiguity in the ground state assignment, excellent agreement is seen between this work, Barker \textit{et al.} \cite{Barker2012}, and theory calculations from Denis \textit{et al.} \cite{Denis2015} and Skripnikov \textit{et al.} \cite{Skripnikov2015}. 

Future directions for a ThF$^+$ eEDM experiment will include ionization spectroscopy of neutral ThF to determine efficient ion creation pathways. The ThF dissociation energy is greater than the ionization energy \cite{Barker2012,Lau1989}, which may allow for greater flexibility in highly-efficient ion creation. One prospect is to prepare the neutral molecule in a core nonpenetrating high-\textit{l} Rydberg state, where it can vibrationally auto-ionize without perturbing the ion core \cite{Eyler1986,Jakubek2001,Kay2008}. Additionally, resonance-enhanced multiphoton dissociation (REMPD) spectroscopy will be performed on ThF$^+$. Laser-induced fluorescence (LIF) spectroscopy had previously been used as the primary detection method on trapped HfF$^+$ \cite{Grau2012}, but REMPD yielded a roughly 100-times improvement in count rate \cite{Ni2014}. We expect to use REMPD as our primary detection scheme in trapped ThF$^+$.

In conclusion, we have demonstrated that comb-vms is a powerful tool for high-resolution extensive survey spectroscopy of molecular ions, in this case ThF$^+$. Our wavelength coverage spans the 696 - 900 nm region, covering 3,260 \cm\ of continuous spectra. Additional spectral regions in the visible and near-IR could be accessed with additional broadening and a different VIPA etalon. 

\textit{Acknowledgements}. Funding was provided by the Marsico Foundation, NIST, and the NSF Physics Frontier Center at JILA (Grant No. 1125844). We thank M. Heaven, A. Titov, L. Skripnikov, A. Petrov, T. Fleig, R. Field, W. Cairncross, and M. Grau for useful discussions. Commercial products referenced in this work are not endorsed by NIST and are for the purposes of technical communication only. 


\bibliographystyle{model1a-num-names}
\bibliography{library}

\onecolumn
\ctable[
caption = 
{\textbf{Fitted constants for observed transitions in ThF$^+$ in cm$^{-1}$.} Quoted uncertainties are statistical and are at the 95\% confidence level. Constants listed are the rotational constant, $B$, centrifugal distortion, $D$, and generic $\Lambda$-doubling constants, $k$, and $k_D$, proportional to $J(J+1)$ and $J^2(J+1)^2$, respectively. Statistical errors on band origins are typically $<$ 90 MHz, but an additional 150 MHz is added to include wavemeter error.} ,
label = transitions, pos = p, 
sideways
]{lcccccccc} {

\tnote[$\circ$]{$k_D'$ unless otherwise noted.}
\tnote[${\dagger}$] {$k_D''$.}
\tnote[$+$] {Values fixed to $B'$ and $D'$ from the $\Omega = 0^+ [10.47]\!\leftarrow\!^3\Delta_1$ (0, 0) transition.}
\tnote[*] {Values fixed to $B'$ and $D'$ from the $\Omega = 0^+ [14.29]\!\leftarrow\!^3\Delta_1$ (0, 1) transition.}
\tnote[$\ddagger$] {Fit to J$_{\text{max}}$ = 50 due to perturbations.}
\tnote[\S]{Fit to J$_{\text{max}}$ = 60.}
}{
	\FL
										&  $\nu_0$ 		&  $B'' $		& $B' $		&  $D'' $		&  $D' $		& $k'' $		& $k' $		& $k_D\ ^\circ$  \NN
										& 				& 			& 			& [$10^{-7}$]	& [$10^{-7}$]	& [$10^{-4}$]	& [$10^{-4}$]	& [$10^{-9}$] \ML
$\Omega = 0^-\!\leftarrow\!^3\Delta_1$ (0, 2)		& 13286.716(7)		& 0.24059(3)	& 0.22898(3)	& 1.29(5)		& 1.34(5)		& -0.887(6)	& \textendash\	& 0 		\NN
$\Omega = 0^-\!\leftarrow\!^3\Delta_1$ (1, 3)		& 13231.099(8)		& 0.23960(4)	& 0.22805(4)	& 1.33(9)		& 1.37(9)		& -0.90(1)		& \textendash\	& 0 		\NN
$\Omega = 0^-\!\leftarrow\!^3\Delta_1$ (0, 1)		& 13935.996(7)		& 0.24161(2)	& 0.22899(2)	& 1.32(2)		& 1.36(2)		& -0.876(3)	& \textendash\	& 0 		 \NN
$\Omega = 0^-\!\leftarrow\!^3\Delta_1$ (1, 2)		& 13876.539(8)		& 0.24061(3)	& 0.22804(3)	& 1.31(5)		& 1.36(5)		& -0.882(6)	& \textendash\	& 0 		 \NN
$\Omega = 0^-\!\leftarrow\!^3\Delta_1$ (2, 3)		& 13817.279(8)		& 0.23962(3)	& 0.22711(3)	& 1.35(3)		& 1.39(3)		& -0.882(5)	& \textendash\	& 0 		 \NN
$\Omega = 0^+ [10.47]\!\leftarrow\!^3\Delta_1$ (0, 0)$^\ddagger$	& 10471.889(7)		& 0.24264(3)	& 0.23534(3)	& 1.30(4)		& 1.33(4)		& 0.869(7)		& \textendash\	& 0 		 \NN
$\Omega = 0^+ [10.47]\!\leftarrow\!^3\Delta_1$ (1, 1)$^\ddagger$	& 10441.560(8)		& 0.24163(6)	& 0.23437(6)	& 1.40(13)		& 1.35(13)		& 0.874(13)	& \textendash\	& 0 		\NN
$\Omega = 0^+ [14.29]\!\leftarrow\!^3\Delta_1$ (0, 1)$^\ddagger$	& 13642.867(7)		& 0.24161(5)	& 0.23194(5)	& 1.37(13)		& 2.33(13)		& 0.892(12)	& \textendash\	&  0 		\NN
$\Omega = 0^+ [14.29]\!\leftarrow\!^3\Delta_1$ (0, 0)$^\ddagger$	& 14295.998(8)		& 0.24259(5)	& 0.23187(5)	& 1.27(15)		& 2.23(15)		& 0.884(14)	& \textendash\	&  0 		\NN
$\Omega = 0^+ [14.29]\!\leftarrow\!^3\Delta_1$ (2, 2)$^\ddagger$	& 13639.836(9)		& 0.24064(7)	& 0.22998(7)	& 1.34(13)		& 1.96(13)		& 0.891(13)	& \textendash\	&  0 		 \NN
$\Omega = 0^+ [10.47]\!\leftarrow\!^1\Sigma^+$ (0, 0)$^\ddagger$	& 10157.607(7)		& 0.24542(3)	& 0.23534(3)$^+$	& 1.35(4)	& 1.33(4)$^+$	& 0		& 0			&  0 		 \NN
$\Omega = 0^+ [14.29]\!\leftarrow\!^1\Sigma^+$ (0, 1)$^\ddagger$	& 13328.320(8)		& 0.24433(3)	& 0.23194(5)*	& 1.46(13)		& 2.33(13)*	& 0			& 0			&  0 		\NN
$\Omega = 0^+ [14.29]\!\leftarrow\!^1\Sigma^+$ (0, 0)$^\ddagger$	& 13981.718(8)		& 0.24545(5)	& 0.23194(5)*	& 1.42(13)		& 2.33(13)*	& 0			& 0			&  0 		\NN
$\Pi_1 [12.99]\!\leftarrow\!^3\Delta_2$ (0, 1)			& 11288.659(7)		& 0.24183(4)	& 0.23001(3)	& 1.34(10)		& 1.4	1(9)		& 0			& 8.71(2)		&  1.3(5) 	\NN
$\Pi_1 [12.99]\!\leftarrow\!^3\Delta_2$ (0, 0)			& 11941.777(6)		& 0.24289(1)	& 0.23005(1)	& 1.37(1)		& 1.44(1)		& 0			& 8.70(1)		&  0.97(7)  \NN
$\Pi_1 [12.99]\!\leftarrow\!^3\Delta_2$ (1, 0)			& 12519.028(7)		& 0.24287(2)	& 0.22918(2)	& 1.35(3)		& 1.43(3)		& 0			& 8.85(1)		&  2.1(2) 	 \NN
$\Pi_1 [12.99]\!\leftarrow\!^3\Delta_2$ (2, 1)$^\S$		& 12433.658(7)		& 0.24179(3)	& 0.22853(3)	& 1.29(6)		& 1.42(6)		& 0			& 9.63(3)		&  7.7(8) 	 \NN
$\Pi_1 [12.99]\!\leftarrow\!^3\Delta_2$ (3, 2)			& 12352.579(7)		& 0.24083(4)	& 0.22775(4)	& 1.37(5)		& 1.44(5)		& 0			& 9.05(3)		&  10.1(7) 	 \NN
$\Pi_1 [14.22]\!\leftarrow\!^3\Delta_2$ (0, 0)			& 13171.134(7)		& 0.24284(4)	& 0.23101(4)	& 1.29(12)		& 1.08(12)		& 0			& -11.84(4)	&  -30(2) 	 \NN
$^3\Pi_2\!\leftarrow\!^3\Delta_3$ (0, 1)$^\ddagger$	& 12877.103(7)		& 0.24236(4)	& 0.22799(4)	& 1.32(11)		& 1.75(11)		& 0			& -0.046(37)	&  74(2) 	 \NN
$^3\Pi_2\!\leftarrow\!^3\Delta_3$ (1, 2)			& 12802.967(8)		& 0.24141(5)	& 0.22675(5)	& 1.46(8)		& 1.42(8)		& 0			& -0.22(3)		&  41(9) 	 \NN
$^3\Pi_2\!\leftarrow\!^3\Delta_3$ (0, 0)$^\ddagger$	& 13532.407(7)		& 0.24337(4)	& 0.22797(4)	& 1.28(17)		& 1.68(17)		& 0			& -0.070(48)	&  74(4) 	 \NN
?	(0, 1)	$^\S$							& 10085.684(7)		& 0.23295(3)	& 0.22322(3)	& 1.37(7)		& 1.31(7)		& 0			& -0.08(2)		& -14.0(7)$^\dagger$ 	\NN
?	(1, 2)									& 10050.200(7)		& 0.23209(2)	& 0.22229(2)	& 1.35(3)		& 1.32(3)		& 0			& -0.09(2)		&  -5.5(3)$^\dagger$  \NN
?	(0, 0)$^\ddagger$						& 10702.830(9)		& 0.23365(3)	& 0.22323(3)	& 1.30(5)		& 1.31(5)		& 0			& -0.19(2)		&  -34.0(7)$^\dagger$  \NN
?	$(v', v'')$								& 12721.532(8)		& 0.23466(5)	& 0.22308(5)	& 1.31(7)		& 1.32(7)		& 0			& 0			&  0 	 \NN
?	$(v'+1, v''+1)$							& 12661.053(8)		& 0.23361(6)	& 0.22203(6)	& 1.16(13)		& 1.17(13)		& 0			& 0			&  0 	\LL
}

\ctable[
caption = {\textbf{Derived constants for observed states in ThF$^+$ in cm$^{-1}$.} T$_0$ is the origin of the $v = 0$ level relative to the $v = 0$ level of $X$ $^3\Delta_1$. T$_0$ errors are directly measured in transitions from $^3\Delta_1$ and $^1\Sigma^+$; otherwise, error bars are adapted from \cite{Barker2012}. Vibrational constants are calculated assuming a Morse potential unless otherwise noted. Quoted errors are 95\%.},
label = stateconstants, pos = hb
]{lcccccccc}{
\tnote[${+}$] {Calculated without assuming a Morse potential.} 
\tnote[*] {$B_0.$}
\tnote[$\dagger$] {Values and error bars from \cite{Barker2012}.} 
\tnote[$\ddagger$] {Calculated using $^3\Delta_2$ and $^3\Delta_3$ energies and errors from \cite{Barker2012}.} 
\tnote[$^\circ$] {Morse potential assumption breaks down for $v' = 2$ and 3.}
}{
	\FL
State				& T$_0$ 			& T$_e$				&  B$_e$ 			&  $\omega_e$ 		& $\omega_e\chi_e$ 	& $\alpha_e$ [$10^{-3}$]	\ML
$^3\Delta_1$		& 0				& 0					& 0.24311(7)		& 656.96(1)$^+$		&  1.920(3)$^+$		& 1.00(4)				\NN
$^1\Sigma^+$	& 314.282(7)			& 314.0(2)				& 0.24601(8)		& 657.90(32)			&  2.26(16)			& 1.12(6)				\NN
$^3\Delta_2$	& 1052.5(1.0)$^\dagger$	& 1052.5(1.0)$^\ddagger$	& 0.24342(2)		& 657.38(22)			&  2.13(10)			& 1.06(4)				\NN
$^3\Delta_3$	& 3150(30)$^\dagger$	& 3149(30)$^\ddagger$	& 0.24388(7)		& 659.31(29)			&  2.00(15)			& 1.01(6)				\NN
$\Omega = 0^+ [10.47]$	& 10471.889(7)		& 10487.1(1)			& 0.23583(5)		& 626.67(56)			&  1.88(17)			& 0.97(7)				\NN
$\Pi_1 [12.99]$		& 12994.3(1.0)$^\ddagger$	& 13032.6(1.0)$^\ddagger$	& 0.23049(2)		& 580.33(10)$^\circ$		&  1.54(5)				& 0.87(2)				\NN
$\Pi_1 [14.22]$		& 14223.6(1.0)$^\ddagger$	& \textendash\			& 0.23101(4)*		& \textendash\			&  \textendash\ 			& \textendash\			\NN
$\Omega = 0^+ [14.29]$	& 14295.967(8)		& \textendash\			& 0.23187(5)*		& \textendash\			&  \textendash\			& \textendash\			\NN
$\Omega = 0^-$	& 14589.09(2)		& 14620.81(1)			& 0.22947(3)		& 593.467(8)$^+$		&  1.822(3)$^+$		& 0.95(4)				\NN
$^3\Pi_2$		& 16682(30)$^\ddagger$	& 16719(30)$^\ddagger$	& 0.22858(7)		& 582.04(75)			&  2.44(18)			& 1.22(6)				\LL
}

\ctable[
caption = {\textbf{Comparison of low-lying states.} Measured T$_e$, B$_e$, and $\omega_e$ values are compared from this work to measured values from Barker \textit{et al.} \cite{Barker2012}, and to theory calculations from Denis \textit{et al.} \cite{Denis2015} and Skripnikov \textit{et al.} \cite{Skripnikov2015}. Quoted uncertainties are 95\%.},
label = theorycomp, pos = p
]{lccccc} {
\tnote[*] {$\Delta G_{1/2}$.}
\tnote[$\dagger$] {Stefan Knecht, unpublished results \cite{Fleig2015}.}
\tnote[$\ddagger$] {Preliminary data for vertical transitions at R$_e$ = 3.75 a.u. \cite{Titov2013}. The accuracy of the calculations has since been improved.}
}{
\FL
	State		&	Constant		&	This Work		&	\cite{Barker2012}	&  \cite{Denis2015}	&  \cite{Skripnikov2015} \ML
$^3\Delta_1$ 	&	T$_e$		&	0			&	315(1)		&	0			&	0 	\NN
			&	B$_e$		&	0.24311(7)	&	0.246(10)		&	0.2439		&	0.2438	\NN
			&	$\omega_e$	&	656.96(1)		&	658.3(2.0)		&	667.3		&	658.4 	\NN
\NN
$^1\Sigma^+$ 	&	T$_e$		&	314.0(2)		&	0			&	318.99		&	448$^\ddagger$ 	\NN
			&	B$_e$		&	0.24601(8)	&	0.245(10)		&	0.2464		&	\textendash	\NN
			&	$\omega_e$	&	657.9(3)		&	656.8(2.0)		&	670.8		&	\textendash 	\NN
\NN
$^3\Delta_2$ 	&	T$_e$		&	\textendash	&	1052.5(1.0)	&	1038.94		&	1186$^\ddagger$ 	\NN
			&	B$_e$		&	0.24342(2)	&	0.243(10)		&	0.2438$^\dagger$	&	\textendash	\NN
			&	$\omega_e$	&	657.38(22)	&	656.5(2.0)		&	667.0$^\dagger$	&	\textendash 	\NN
\NN
$^3\Delta_3$ 	&	T$_e$		&	\textendash 	&	3150(30)		&	3161.99		&	3193$^\ddagger$ 	\NN
			&	B$_e$		&	0.24388(7)	&	\textendash	&	0.2443$^\dagger$	&	\textendash	\NN
			&	$\omega_e$	&	659.33(26)	&	665(30)*		&	669.0$^\dagger$	&	\textendash 	\LL		
}

\ctable[
caption = 
{\textbf{Fitted constants for observed transitions in ThO$^+$.} Errors are statistical only and at the 95\% confidence level. Bands were fit with half-integer J. No $\Lambda$-doubling terms were measured. Statistical errors on band origins are $<$ 60 MHz, but an additional 150 MHz is added to include wavemeter error. No electronic states were able to be conclusively assigned.} ,
label = tho, pos = p]{lccccc} {

}{
	\FL
					&  $\nu_0$ 			&  $B'' $		& $B' $		&  $D'' $		&  $D' $		\NN
					& 					& 			& 			& [$10^{-7}$]	& [$10^{-7}$]	\ML
?	(0, 0)				& 12146.812(6)			& 0.33984(2)	& 0.32610(1)	& 1.91(2)		& 1.89(2)		\NN
?	(1, 1)				& 12091.106(7)			& 0.33843(4)	& 0.32469(4)	& 1.78(7)		& 1.77(6)		\NN
?	(1, 0)				& 13006.858(7)			& 0.33973(4)	& 0.32465(4)	& 1.6(1)		& 1.6(1)		\NN
?	(0, 0)				& 11564.329(6)			& 0.33900(2)	& 0.32612(2)	& 1.86(2)		& 1.87(2)		\NN
?	(1, 1)				& 11510.656(7)			& 0.33767(3)	& 0.32480(3)	& 1.86(5)		& 1.87(5)		\LL
}	

\end{document}